\documentclass{nature}

\usepackage{lettrine}
\usepackage{dcolumn}
\usepackage{bm}
\usepackage{amssymb}
\usepackage{epsfig}
\usepackage[10pt]{moresize}
\usepackage{comment}
\usepackage{epsfig}
\usepackage{subfigure}
\usepackage{graphicx,morefloats,float}
\usepackage{color}
\usepackage{amsmath}
\usepackage{bbm}

\title{Single-step transfer or exchange of multipartite quantum entanglement with minimum resources}
\author{Chui-Ping Yang$^{1}$, Qi-Ping Su$^{1}$, Shi-Biao Zheng$^{2\star}$, and Siyuan Han$%
^{3,4\star}$}
\begin{document}
\maketitle
\begin{affiliations}
\item Department of Physics, Hangzhou Normal University, Hangzhou, Zhejiang 310036, China
\item Department of Physics, Fuzhou University, Fuzhou 350002, China
\item Department of Physics and Astronomy, University of Kansas, Lawrence, Kansas 66045, USA
\item Beijing National Laboratory for Condensed Matter Physics,
Institute of Physics, Chinese Academy of Sciences, Beijing 100190, China\\
$^\star$ Correspondence and requests for materials should be addressed to S.Han (email: han@ku.edu) or S.B.Zheng (sbzheng11@163.com)
\end{affiliations}

\begin{abstract}
The transfer or exchange of multipartite quantum
states is critical to the realization of large-scale quantum information
processing and quantum communication. A challenging question in this context
is: \textquotedblleft What is the minimum resource required and how to
simultaneously transfer or exchange multipartite quantum entanglement
between two sets of qubits". Finding the answer to these questions is of
great importance to quantum information science. In this work, we
demonstrate that by using a single quantum two-level system - the simplest
quantum object - as a coupler arbitrary multipartite quantum states (either
entangled or separable) can be transferred or exchanged simultaneously
between two sets of qubits. Our findings offer the potential to
significantly reduce the resources needed to construct and operate
large-scale quantum information networks consisting of many multi-qubit
registers, memory cells, and processing units.
\end{abstract}

\newpage Entanglement arises from nonclassical correlation between the
constituents of multipartite quantum systems. It is one of the most profound
and difficult to understand aspects of quantum physics. Entanglement is
indispensable in quantum information science as demonstrated by Shor's
factorization algorithm [1] and various quantum key distribution protocols
[2,3]. Recently, considerable interest has been devoted to the application
of entangled states in quantum computation [4,5], quantum cryptography
[2,6], teleportation [7-9], and quantum copying [10,11] and many previously
unknown or unexpected properties of entanglement, such as entanglement
swapping [10] and entanglement sudden death [12], have been discovered. Over
the past decade, experimentalists have generated and verified entanglement
in a variety of physical systems, including eight photons via linear optical
devices [13,14], fourteen trapped ions [15], two atoms in cavity QED
[16,17], two excitons in a single quantum dot [18], electron spins in two
proximal nitrogen-vacancy centres [19], and up to five superconducting
qubits coupled via a single cavity or capacitors [20-25].

Because transfer or exchange of arbitrary multipartite states (TEAMS) is of
great importance to utilizing entanglement for quantum information
processing (QIP) and quantum communication, it has attracted much attention.
In principle, TEAMS can be accomplished by expanding either
entanglement-based quantum teleportation protocols or non-teleportation
protocols. For instance, many theoretical schemes [26-30] and experiments
[31-35] have investigated how to transfer or exchange quantum states between
two qubits using \textit{entanglement-based} quantum teleportation protocols
[7]. \ Among experiments, quantum state transfer between two superconducting
qubits has been demonstrated in circuits consisting of multiple
superconducting qubits coupled to planar resonators [36-39]. \
Alternatively, quantum state transfer or exchange can also be realized using
non-teleportation protocols. For instance, by using photons (transmitted via
an optical fiber) as the information carriers the transfer of quantum states
from one atom to another has been explored [40-42]. In addition, a quantum
network, with single atoms placed in fiber-connected cavities, has been
proposed and the transfer of atomic quantum states and the creation of
entanglement between two distant nodes of the network have been demonstrated
experimentally [43].

Because in the work mentioned above the states being transferred or
exchanged are single particle states, it is not granted that these protocols
can be applied to multipartite states without a substantial increase of
resources (e.g., multiple EPR pairs). \ As quantum networks play an
increasingly important role in scalable QIP it is imperative to explore new
methods to realizing TEAMS with a minimum amount of resources. \ Some of the
most urgent issues in this context include \textquotedblright What is the
minimum quantum hardware resource (e.g., the number of qubits and couplers)
required to transfer or exchange arbitrary multipartite quantum states
between two quantum registers each having $N$ qubits?\textquotedblright\ and
\textquotedblright Given the minimum quantum hardware resource, could
transfer or exchange of $N$-partite states be done with a single step of
operation?\textquotedblright . Positive answers to these questions would not
only have significant impact on the architecture of future quantum networks
but also is of highly interesting to the foundation of quantum mechanics.

In this work, we answer these two critical questions by considering a
generic model system consisting of $2N$\ qubits (e.g., spin $\frac{1}{2}$\
particles) coupled to a two-level coupler C (Fig. 1). \ The $2N$\ qubits are
divided arbitrarily into two sets, labelled as the set A and set B
respectively, each containing $N$\ qubits. \ It is also assumed that qubits
in the same set may or may not have direct intra-set coupling and that no%
\textrm{\ }\textit{direct} coupling exists between qubits in different sets.
The two-level coupler acts as an intermediary to allow quantum information,
in the form of multipartite quantum states, flow from A to B and vice versa.
\ We show that for $N\geq 2$ by multiplexing a \textit{single two-level
coupler} is sufficient to generate coupler-mediated effective interaction
between the $N$ pairs of qubits and that arbitrary\textrm{\ }$N$-partite
states can be transferred or exchanged between A and B in a \textit{single}
step\textrm{. }Namely, the \textit{minimum} quantum hardware resource to
transfer or exchange a piece of $N$-bit quantum information simultaneously
in one step between two sets is a single quantum two-level coupler. In
addition, the coupler can also be used to mediate interactions between
qubits in the same set, allowing creation and manipulation of entanglement
within each set.

\bibliographystyle{naturemag}
\bibliography{ms.bib}

This result is nontrivial and not known \textit{a priori} because the
Hilbert space of each $N$-qubit set is\textrm{\ }$2^{N}$-dimensional whereas
that of the coupler is only $2$-dimensional, which is the minimum for any
quantum systems, and the states to be exchanged or transferred between A and
B registers are arbitrary $N$-partite states (e.g., entangled or separable,
pure or mixed). \ According to conventional wisdoms, one would think that
transferring or exchanging quantum information, which requires a $2^{N}$%
-dimensional Hilbert space to accommodate, between two $N$-qubit sets in one
step via a single coupler would require the coupler also having at least a $%
2^{N}$-dimensional Hilbert space. Thus, it is natural to think that
transferring or exchanging $N$-qubit states would require $N$ auxiliary
two-level coupler (TLC) plus one operational step, or alternatively one TLC
plus $N$ repeated operational steps, to accomplish.

We point out that the method proposed here has several distinctive
advantages: (i) Only a two-level coupler is needed, and TEAMS can be
performed simultaneously in a single step without the use of classical
rf/microwave/optical pulses during the state transfer/exchange operation.
This unique feature reduces the complexity of the circuits and operations.
(ii) The two-level coupler C can be either a true quantum two-level system
(TLS), such as an electron spin, or an effective TLS, such as the two lowest
levels of a superconducting qubit, so that the scheme can be applied to a
large variety of physical quantum information networks. (iii) During the
operation the coupler stays mostly in its ground state so that the effects
of quantum channel decoherence is greatly suppressed. This property allows
the use of couplers with shorter decoherence time but has other desirable
attributes such as rapid frequency tunability, design flexibility, or good
scalability. (iv) It offers the flexibility of reconfiguring interactions
between pairs of qubits, either intra-set or inter-set, \textit{in situ} to
perform various QIP tasks without changing hardware wirings. (v) By
connecting the qubits to multiple, e.g., two or three two-level couplers,
the structure can be expanded into one- or two-dimensional quantum networks
- a promising architecture for scalable QIP.

In what follows, we derive the interaction Hamiltonian that governs the
system dynamics of the $2N$ qubits plus one two-level coupler. It is evident
from the Hamiltonian that $N$ pairs of \textit{in situ} programmable
qubit-qubit superexchange interaction can occur in parallel without
interference to each other\ allowing the possibility of realizing TEAMS in a
single step (e.g., by making all coupler-mediated effective pair
interactions the same strength). As an example, we describe in detail how to
perform $N$-partite state exchange (swap) and transfer using this generic
configuration. Furthermore, we propose a circuit QED-based implementation of
the scheme. With realistic device and circuit parameters, numerical
simulations show that\ the fidelity can reach $99.1\%$ for Bell-state
transfer and no less than $96.3\%$ for Bell-state swap. Finally, we
summarize the key result and its impact on the future development of quantum
information science.

\section*{Results}

\noindent \textbf{Hamiltonian.} Without the loss of generality we consider
two sets of otherwise noninteracting qubits connected to a two-level coupler
C, hereafter referred to as coupler C for simplicity, as illustrated in
Fig.~1(a). The first set contains $N$ qubits $\left\{
a_1,a_2,...a_j,...a_N\right\} $ while the second set contains the remaining $%
N$ qubits $\left\{ b_1,b_2,...,b_k,...b_N\right\} $. The two logic states of
the qubits $a_j$ ($b_k)$ are labelled as $\left| 0\right\rangle _{a_j(b_k)}$
and $\left| 1\right\rangle _{a_j(b_k)}$ and that of the coupler C are
denoted as $\left| g\right\rangle _{\mathrm{c}}$ and $\left| e\right\rangle
_{\mathrm{c}},$ respectively. For qubit $a_j,$ we define the operators $\hat
a_j$ and $\hat a_j^{+},$ which satisfy $\hat a_j\left| 0\right\rangle
_{a_j}=0,$ $\hat a_j\left| 1\right\rangle _{a_j}=\left| 0\right\rangle
_{a_j},$ and $\hat a_j^{+}\left| 0\right\rangle _{a_j}=\left| 1\right\rangle
_{a_j}.$ By replacing \textquotedblleft $a_j"$\ by \textquotedblleft $b_k"$\
the operators $\hat b_k$\ and $\hat b_k^{+}$\ are defined for qubit $b_k$.
In addition, we define the raising and\textrm{\ }lowering\textrm{\ }%
operators $\sigma =\left| g\right\rangle _{\mathrm{c}}\left\langle e\right| $
and $\sigma ^{+}=\left| e\right\rangle _{\mathrm{c}}\left\langle g\right| $
for the coupler C, which satisfy $[\sigma ^{+},\sigma ]=\sigma _z$ with $%
\sigma _z=\left| e\right\rangle _{\mathrm{c}}\left\langle e\right| -\left|
g\right\rangle _{\mathrm{c}}\left\langle g\right| .$ The discussion below is
based on Fig.~1(a). However, it should be mentioned that the results can
directly apply to Fig.~1(b) to accomplish the same tasks, by mapping\textrm{%
\ }the large detuning conditions, required for the qubit pairs ($a_1,b_1$), (%
$a_2,b_2$),..., and ($a_N,b_N$), to the qubit pairs ($a_1,b_2$), ($a_2,b_N$%
),..., and ($a_N,b_1$) in Fig.~1(b), respectively.

In general, qubits $a_j$\ and $b_k$\ can be tuned to have the same detuning\
with respect to the coupler's transition frequency $\omega _C.$\ However,
for the sake of simplicity, we set $j=k$\ in the following discussion.
Suppose qubit $a_j$ ($b_j$) is coupled to the coupler C, with coupling
strength $g_j$ ($\mu _j$) and detuning\ $\Delta _j$. In the interaction
picture, the Hamiltonian of the whole system is given by
\begin{equation}
H_I=\sum_{j=1}^N\left( g_je^{i\Delta _jt}\hat a_j\sigma ^{+}+\mu
_je^{i\Delta _jt}\hat b_j\sigma ^{+}+\mathrm{H.c.}\right) ,
\end{equation}
where $\Delta _j=\omega _{\mathrm{c}}-\omega _{aj}=\omega _{\mathrm{c}%
}-\omega _{bj}$ (Fig. 2) and $\omega _{aj}$ ($\omega _{bj}$) is the
frequency of qubit $a_j$ ($b_j$).

Under the large detuning condition $\Delta _j\gg g_j,\mu _j,$\ the two sets
of qubits do not exchange energy with the coupler. However, the coupler can
mediate $N$ independent pair-wise superexchange interactions between the two
sets of $2N$ qubits. Qubit $a_j$\ is only coupled to qubit $b_j$\ when the
detunings satisfy the following conditions
\begin{equation}
\frac{\left| \Delta _j-\Delta _k\right| }{\Delta _j^{-1}+\Delta _k^{-1}}%
>>g_jg_k,\;g_j\mu _k,\;\mu _j\mu _k;\;j\neq k.
\end{equation}
Then we obtain the effective Hamiltonian $H_{eff}=H_0+H_{int},$ with
\begin{eqnarray}
H_0 &=&\sum_{j=1}^N\left( \frac{g_j^2}{\Delta _j}\hat a_j\hat a_j^{\dagger }+%
\frac{\mu _j^2}{\Delta _j}\hat b_j\hat b_j^{\dagger }\right) \left|
e\right\rangle \left\langle e\right|  \notag \\
&&-\sum_{j=1}^N\left( \frac{g_j^2}{\Delta _j}\hat a_j^{\dagger }\hat a_j+%
\frac{\mu _j^2}{\Delta _j}\hat b_j^{\dagger }\hat b_j\right) \left|
g\right\rangle \left\langle g\right| ,
\end{eqnarray}
\begin{equation}
H_{int}=\sum_{j=1}^N\lambda _j(\hat a_j\hat b_j^{\dagger }+\hat a_j^{\dagger
}\hat b_j)(\left| e\right\rangle \left\langle e\right| -\left|
g\right\rangle \left\langle g\right| ),
\end{equation}
where $\lambda _j=g_j\mu _j/\Delta _j.$\ The first (second) term in the
first bracket of $H_0$\ is an ac-Stark shift of the level $\left|
e\right\rangle $\ of the coupler C, induced by the interaction with qubit $%
a_j$\ ($b_j$); while the first (second) term in the second bracket of $H_0$\
is an ac-Stark shift of the level $\left| g\right\rangle $\ of the two-level
coupler, induced by the interaction with qubit $a_j$\ ($b_j$). Here and
below, we have defined $\left| g\right\rangle \equiv $\ $\left|
g\right\rangle _{\mathrm{c}}$\ and $\left| e\right\rangle \equiv $\ $\left|
e\right\rangle _{\mathrm{c}}$ for simplicity.

To simplify discussions hereafter we set $g_{j}=\mu _{j}$ and $\omega
_{aj}=\omega _{bj}=\omega _{j}$ which can be realized readily by design and
fabrication. Consequently, the qubits $a_{j}$\ and $b_{j}$\ have the same
detuning $\Delta _{j}$. It is also understood that $\omega _{i}\neq \omega
_{j}$\ and $g_{i}\neq g_{j}$\ for $i\neq j.$\ In this way, each pair of
qubits has its own unique frequency and qubit-coupler interaction strength
while all pairs have the same effective coupler mediated interaction
strength.\emph{\ }In a new interaction picture\emph{\ }with respect to the
Hamiltonian $H_{0},$ we have $H_{int}^{\prime
}=e^{iH_{0}t}H_{int}e^{-iH_{0}t}=H_{int}.$ When the coupler C is initially
in the ground state $\left\vert g\right\rangle $, it will remain in this
state throughout the interaction as the Hamiltonian $H_{int}$\ cannot induce
any transition for the coupler. In this case, based on Eq. (4) and $%
H_{int}^{\prime }=H_{int},$\ the Hamiltonian $H_{int}^{\prime }$\ is reduced
to\textrm{\ }

\begin{equation}
H_e=-\sum_{j=1}^N\lambda _j(\hat a_j\hat b_j^{\dagger }+\hat a_j^{\dagger
}\hat b_j),
\end{equation}
which is the effective Hamiltonian governing the dynamics of the two sets of
qubits.

The two sets of qubits can be any type of qubits such as bosonic qubits or
atomic qubits (e.g., artificial atoms or natural atoms). In principle, we
can employ this effective Hamiltonian to implement several fundamental
quantum operations on two sets of qubits, such as\ entanglement swap,
multi-qubit logic gates, and creation of quantum entanglement in or\textrm{\
}between two sets of qubits. As a concrete example, in the next section we
explicitly show how to apply this Hamiltonian to implement TEAMS between two
sets of bosonic qubits.

As a final note, we point out that the condition $g_{j}=\mu _{j}$ is
unnecessary. As shown in the Method, for the case of $g_{j}\neq \mu _{j}$,
the effective Hamiltonian (5) can be obtained by setting the detuning of the
qubit $a_{j}$ slightly different from that of qubit $b_{j}$ ($j=1,2,...,N$).

\noindent \textbf{Quantum state swapping and transfer.} Let us go back to
Fig. 1(a), where any initially unentangled state of the first set of $N$
qubits $(a_1,a_2,...,a_N)$ and the second set of $N$ bosonic qubits ($%
b_1,b_2,...,b_N$) can be described by the joint state $\left| \psi _{\mathrm{%
A}}\left( 0\right) \right\rangle \otimes \left| \psi _{\mathrm{B}}\left(
0\right) \right\rangle .$ Here, the first (second)\ part of the product is
the initial state of the first (second) set of $N$ qubits, taking a general
form of $\left| \psi _{\mathrm{A}}\left( 0\right) \right\rangle
=\sum_{n_j=0}^1c_{\{n_j\}}\prod_{j=1}^N\left| n_j\right\rangle _{a_j}$ ($%
\left| \psi _{\mathrm{B}}\left( 0\right) \right\rangle
=\sum_{m_k=0}^1d_{\{m_k\}}\prod_{k=1}^N\left| m_k\right\rangle _{b_k}$). The
subscript $a_j$\ ($b_k$) represents qubit $a_j$\ ($b_k$), $c_{\{n_j\}}$\ is
the coefficient of the component $\prod_{j=1}^N\left| n_j\right\rangle
_{a_j} $\ of the initial state for the qubits ($a_1,a_2,...,a_N$), and the
same notation applies to $d_{\{m_k\}}$\ for the qubits ($b_1,b_2,...,b_N$).
In terms of $\left| 1_j\right\rangle _{a_j}=\hat a_j^{\dagger }\left|
0\right\rangle _{a_j}$ and $\left| 1_k\right\rangle _{b_k}=\hat b_k^{\dagger
}\left| 0\right\rangle _{b_k},$ we can write down the initial state as
\begin{eqnarray}
&&\ \ \ \left| \psi _{\mathrm{A}}\left( 0\right) \right\rangle \otimes
\left| \psi _{\mathrm{B}}\left( 0\right) \right\rangle  \notag \\
\
&=&\sum_{n_j=0,1}c_{\{n_j\}}\sum_{m_k=0,1}d_{\{m_k\}}\prod_{j=1}^N%
\prod_{k=1}^N\left( \hat a_j^{+n_j}\hat b_k^{+m_k}\left| 0\right\rangle
_a\left| 0\right\rangle _b\right) ,
\end{eqnarray}
where $\left| 0\right\rangle _a=\left| 0\right\rangle _{a_1}...\left|
0\right\rangle _{a_N}$ and $\left| 0\right\rangle _b=\left| 0\right\rangle
_{b_1}...\left| 0\right\rangle _{b_N}.$

For bosonic qubits, the operators ($\hat{a}_{j},\hat{a}_{j}^{+}$) and ($\hat{%
b}_{j},\hat{b}_{j}^{+}$) obey $\left[ \hat{a}_{j},\hat{a}_{j}^{+}\right] =$ $%
\left[ \hat{b}_{j},\hat{b}_{j}^{+}\right] =1.$ The effective Hamiltonian $%
H_{e}$ leads to the transformations $e^{-iH_{e}t}\hat{a}_{j}^{\dagger
}e^{iH_{e}t}=\cos (\lambda _{j}t)\hat{a}_{j}^{\dagger }+i\sin (\lambda _{j}t)%
\hat{b}_{j}^{\dagger },$ and $e^{-iH_{e}t}\hat{b}_{j}^{\dagger
}e^{iH_{e}t}=\cos (\lambda _{j}t)\hat{b}_{j}^{\dagger }+i\sin (\lambda _{j})%
\hat{a}_{j}^{\dagger }.$ These transformations have the following property:
(i) by setting $\left\vert \lambda _{j}\right\vert =\lambda ,$ i.e., $%
g_{j}\mu _{j}/\left\vert \Delta _{j}\right\vert =\lambda $ (independent of $j
$). This condition can be met by using frequency-tunable qubits (or
resonators). In the case of fixed frequency resonators one can design and
fabricate the qubits $a_{j}$\ and $b_{j}$\ to have the proper frequencies ($%
\omega _{aj}=\omega _{bj}=\omega _{j}$) and coupling strengths\  ($g_{j},$ $%
\mu _{j}$) respectively and to set $\left\vert \Delta _{j}\right\vert
=g_{j}\mu _{j}/\lambda $ accordingly, and (ii) for $\lambda t=\pi /2,$ we
obtain $e^{-iH_{e}t}\hat{a}_{j}^{\dagger }e^{iH_{e}t}=i^{\lambda
_{j}/\lambda }\hat{b}_{j}^{\dagger }$ and $e^{-iH_{e}t}\hat{b}_{j}^{\dagger
}e^{iH_{e}t}=i^{\lambda _{j}/\lambda }\hat{a}_{j}^{\dagger }.$ Accordingly,
we have\emph{\ }$e^{-iH_{e}t}\hat{a}_{j}e^{iH_{e}t}=-i^{\lambda _{j}/\lambda
}\hat{b}_{j}$ and $e^{-iH_{e}t}\hat{b}_{j}e^{iH_{e}t}=-i^{\lambda
_{j}/\lambda }\hat{a}_{j}.$ These unitary transformations will be employed
in the derivation of Eq. (7) below.

Under the Hamiltonian $H_e,$ the state of the subsystem, consisting of the $%
2N$ qubits in sets A and B, after an evolution time $t=\pi /\left( 2\lambda
\right) $ is given by
\begin{eqnarray}
\left| \psi _{\mathrm{AB}}\left( t\right) \right\rangle &=&e^{-iH_et}\
\left| \psi _{\mathrm{A}}\left( 0\right) \right\rangle \otimes \left| \psi _{%
\mathrm{B}}\left( 0\right) \right\rangle  \notag \\
\
&=&\sum_{n_j=0,1}c_{\{n_j\}}\sum_{m_k=0,1}d_{\{m_k\}}\prod_{j=1}^N%
\prod_{k=1}^N  \notag \\
&&\left[ \left( i\right) ^{n_j\lambda _j/\lambda }\left( i\right)
^{m_k\lambda _k/\lambda }\left( \hat b_j^{\dagger }\right) ^{n_j}\left( \hat
a_k^{\dagger }\right) ^{m_k}\left| 0\right\rangle _a\left| 0\right\rangle _b%
\right]  \notag \\
\ &=&\sum_{m_k=0,1}d_{\{m_k\}}\prod_{k=1}^N\left( i\right) ^{m_k\lambda
_k/\lambda }\left| m_k\right\rangle _{a_k}  \notag \\
&&\otimes \sum_{n_j=0,1}c_{\{n_j\}}\prod_{j=1}^N\left( i\right) ^{n_j\lambda
_j/\lambda }\left| n_j\right\rangle _{b_j},
\end{eqnarray}
where $\lambda _j/\lambda =\pm 1$ and $\lambda _k/\lambda =\pm 1.$ Note that
in the last two lines of Eq. (7), the first part of the product represents
the $N$-qubit state of ($a_1,a_2,...,a_N$) while the second part is that of (%
$b_1,b_2,...,b_N$).

After returning to the original interaction picture, the state of the whole
system, $\left\vert \psi _{\mathrm{ABC}}^{\prime }\left( t\right)
\right\rangle =e^{-iH_{0}t}\left\vert \psi _{\mathrm{AB}}\left( t\right)
\right\rangle \left\vert \psi _{\mathrm{c}}\left( t\right) \right\rangle ,$
can be further written as $\left\vert \psi _{\mathrm{ABC}}^{\prime }\left(
t\right) \right\rangle =\left\vert \psi _{\mathrm{AB}}^{\prime }\left(
t\right) \right\rangle \otimes \left\vert g\right\rangle _{\mathrm{c}}.$ By
letting $H_{0}$ act on the state $\left\vert \psi _{\mathrm{AB}}\left(
t\right) \right\rangle ,$ we obtain a decomposition of $\left\vert \psi _{%
\mathrm{AB}}^{\prime }\left( t\right) \right\rangle =\left\vert \psi _{%
\mathrm{A}}\left( t\right) \right\rangle \otimes \left\vert \psi _{\mathrm{B}%
}\left( t\right) \right\rangle $ with
\begin{equation}
\left\vert \psi _{\mathrm{A}}\left( t\right) \right\rangle
=\sum_{m_{k}=0,1}d_{\{m_{k}\}}\prod_{k=1}^{N}\left( e^{i\phi _{k}m_{k}\pi
}\left\vert m_{k}\right\rangle _{a_{k}}\right) ,
\end{equation}%
\begin{equation}
\left\vert \psi _{\mathrm{B}}\left( t\right) \right\rangle
=\sum_{n_{j}=0,1}c_{\{n_{j}\}}\prod_{j=1}^{N}\left( e^{i\theta _{j}n_{j}\pi
}\left\vert n_{j}\right\rangle _{b_{j}}\right) ,
\end{equation}%
where $\phi _{k}=(\lambda _{k}+g_{k}^{2}/\Delta _{k})/(2\lambda )$ and $%
\theta _{j}=(\lambda _{j}+\mu _{j}^{2}/\Delta _{j})/(2\lambda )$. This is
equivalent to the quantum state swap operation plus single-qubit phase
shifts $e^{i\phi _{k}\pi }$ ($e^{i\beta _{j}\pi }$) on the state $\left\vert
1\right\rangle $ of qubit $a_{k}$ ($b_{j}$). These additional phase shifts
can be corrected for by local single-qubit rotations $e^{-i\phi _{k}\pi \hat{%
a}_{k}^{\dagger }\hat{a}_{k}}$ and $e^{-i\theta _{j}\pi \hat{b}_{j}^{\dagger
}\hat{b}_{j}}$. Notice that the multiplexed quantum state exchange protocol
described above becomes the state transfer protocol by initializing all
qubits in the second (i.e., receiving) set in the state $\left\vert
0\right\rangle $. More importantly, because the states $\left\vert \psi _{%
\mathrm{A}}\left( 0\right) \right\rangle $ and $\left\vert \psi _{\mathrm{B}%
}\left( 0\right) \right\rangle $ considered above take a general form, the
protocol can be applied directly to swap or transfer any type of
multipartite entanglement, such as the GHZ state $\left\vert
00...0\right\rangle +\left\vert 11...1\right\rangle ,$ the W- state $\frac{1%
}{\sqrt{N}}\left( \left\vert 00...001\right\rangle +\left\vert
00...010\right\rangle +...+\left\vert 10...000\right\rangle \right) ,$ the
cluster state, and so on, between the two sets of multiple qubits.

It should be mentioned that in reality a physical coupler usually has more
than two levels. However, if the coupler is a nonlinear quantum element such
as a superconducting qubit, population leakage out of the two-dimensional
Hilbert space formed by $\left\vert g\right\rangle $ and $\left\vert
e\right\rangle $ of the coupler can be made negligible by choosing proper
coupler parameters. In contrast, when the coupler\ is a single-mode
resonator [44], the probability of population leaking into higher energy
levels of the coupler could be significant due to its uniform energy level
spacing. This problem becomes apparent as the number of qubits increases.

Quantum dynamics of two bosonic qubits/resonators coupled by a
superconducting qubit as a quantum switch has been studied previously in
[45,46]. However, although our method of TEAMS is based on the same type of
coupler mediated dispersive interaction between qubits described in [45,46]
it is not a simple extension of the latter because that would require the
use of $N$ couplers for $N$ pairs of qubits/resonators. The distinctive
feature of our method is to utilize the "frequency multiplexing" capability
of our effective Hamiltonian to have each qubit\ in one set coupled uniquely
to only one of the qubits in the other set\ and to have all $N$ pair-wise
interactions occur concurrently, so that one-step TEAMS between the two $N$%
-qubit sets with only one coupler qubit, rather than $N$\ couplers, becomes
possible.

It is noted that if one chooses to perform TEAMS between two sets of
resonators the preparation of the initial state of the resonators would in
general require the use of qubits as well as tunable qubit-resonator
couplings [47-50]. For example, this task could be accomplished by coupling
one ancilla qubit to each resonator [51,52]. However, because the main
objective of this work is to show how to perform TEAMS in a single step we
assume the states to be transfered or exchanged already exist. Thus, we will
not discuss the details of how to prepare the initial states of the
resonators.

The TLC is assumed to be a frequency-tunable superconducting
qubit (a.k.a. artificial atom) [53-56]. Generally speaking, it is highly
desirable to use qubits with frequency and coupling strength ($g_{j}$\ and $%
\mu _{j})$\ both tunable to implement the proposed one-step TEAMS as the
double tunability would provide great flexibility in satisfying all required
conditions, in particular $|\lambda _{j}|=g_{j}\mu _{j}/|\Delta
_{j}|=\lambda $. In practice, however, frequency tunability is readily
available for artificial atoms and to a less extent for resonators [57,58]
while tunable coupling strength is significantly more difficult to obtain.

We emphasize that assumption of uniform\ effective coupling strength is
unnecessary and it is only used for the convenience of discussion above. For
instance, a manufactured circuit with fixed coupling strengths may have $j$%
-denpendent\ effective coupling strengths $\lambda _{j}$. In this case, the
TEAMS cannot be completed by turning on/off the effective coupling for all
pairs of qubits simultaneously. Fortunately, this problem can be
circumvented by relaxing the strong condition to a weaker one: instead to
require all $\lambda _{j}$'s to have the same magnitude they can be
different as long as the condition $\omega _{aj}=$\ $\omega _{bj}=\omega
_{j}\neq \omega _{i\neq j}$\ is still satisfied. The weaker condition can be
met by using frequency tunable qubits or resonators. A simple case to
consider is the qubit-coupler coupling strengths for all $2N$\ qubits
(resonators) are the same or approximately equal. \ Experimentally, this is
the easiest to realize and most likely to be encountered. With this set up
all one needs to do is to switch on the effective dispersive interaction
between qubits $a_{j}$\ and $b_{j}$\ at a proper time $\tau _{j}=t_{\max
}-t_{j}$\ by tuning their frequencies to have the proper $\Delta _{j},$\
where $t_{\max }=\max (\pi /2\lambda _{1},$\ $\pi /2\lambda _{2},...$\ $\pi
/2\lambda _{N})$\ and $t_{j}=\pi /2\lambda _{j},$\ and let it evolve for a
time interval $t_{j}$\ before switching off the effective interaction $%
\lambda _{j}$. Consequently, at $t=t_{\max }$\ all coupler mediated
effective interactions are switched off which can be accomplished by tuning
the $coupler$\ frequency $\omega _{c}$ far way from that of all $2N$\
qubits. In this last step the coupler is used essentially as a quantum
switch [45,46] to simultaneously cut off the effective interaction between
all pairs of qubits.

The coupling between the resonators and the coupler qubit can be effectively
turned on (off) by adjusting the level spacings of the coupler qubit. When
the coupler qubit frequency is highly detuned from the resonator frequencies
the couplings are effectively switched off, and when the coupler qubit
frequency is detuned from the resonator frequencies by a suitable amount
they are dispersively coupled as the case discussed above. For a
superconducting coupler qubit, the level spacings can be readily adjusted by
varying external control parameters (e.g., magnetic flux applied to phase,
transmon, or flux qubits, see, e.g., [53-56]).

\noindent \textbf{Experimental implementation. }In practice, the proposed
scheme can be implemented using either the artificial atoms (e.g.,
superconducting qubits) or resonators [e.g., superconducting co-planar
waveguide (CPW) resonators] as the physical objects to demonstrate the
proposed one-step TEAMS protocol. The artificial atoms have the advantage
of\ tunnable frequency, better separation between the computational states
from the non-computational ones because they are nonlinear oscillators, and
the ease of initial state preparation. On the other hand, high-$Q$\ CPW
resonator is comparatively easier to design and fabricate. For example, CPW
resonators with quality factor on the order of $10^{6}$\ (i.e., about $30$ $%
\mu $s of the lifetime of photons for a $6$ GHz resonator) have been
demonstrated with a single layer of sputtered superconducting films [59-61].
\ In addition, frequency tunnable resonators have also been demonstrated
recently [57,58].\textrm{\ }

In the example discussed below, we choose resonators as the realization of
bosonic qubits for the following reasons: (1) Systems of superconducting
resonators and qubits have been considered one of the most promising
candidates for quantum information processing [62-65] and there is a growing
interest in quantum information processing based on microwave photon
qubits.\ Within circuit QED, several theoretical proposals have been put
forward for utilizing microwave photons stored in two superconducting CPW
resonators as qubits/qudits for quantum gates [66-69]. (2) Microwave photons
have been considered as candidates for quantum memories [58,70-72]. When
performing quantum information processing, TEAMS between different
multi-qubit memory banks would become a ubiquitous task. (3) Because it is
in general more difficult to tune the frequency of the resonantors than
artificial atoms and linear resonators are a poor realization of qubits, if
our scheme can be demonstrated to work well with frequency and coupling
strength non-tunable resonators it would work better and/or easier to
implement with frequency tunable artificial atoms or resonators. Namely, we
choose a more difficult case to study.

Let us now consider four fixed-frequency superconducting coplanar waveguide
(CPW) resonators, capacitively coupled to a superconducting transmon coupler
[73] as illustrated in Fig. 3. We emphasize again that using frequency
tunable resonators would make the implementation considerably easier. For
simplicity, we use ($a_{1},a_{2},b_{1},b_{2}$) to denote the four qubits.
For the setup here, $a_{j}$ ($b_{j}$) is a bosonic mode\emph{\ }of the
resonator $a_{j}$ ($b_{j}$), and the two logic states of the qubit $a_{j}$ ($%
b_{j}$) are represented by the vacuum state and the single-photon state of
the bosonic mode of resonators $a_{j}$ ($b_{j}$) ($j=1,2$). In the
following, we first present a general discussion on the fidelity of the
operation. To quantify operation fidelity of the proposed protocol, we then
numerically calculate the fidelity for transferring and exchanging each of
the four Bell states $\left\vert \psi ^{\pm }\right\rangle =\frac{1}{\sqrt{2}%
}\left( \left\vert 01\right\rangle \pm \left\vert 10\right\rangle \right) $
and $\left\vert \phi ^{\pm }\right\rangle =\frac{1}{\sqrt{2}}\left(
\left\vert 00\right\rangle \pm \left\vert 11\right\rangle \right) $ between
the two pairs of qubits (i.e., the case of $N=2$).

In the above discussions, we have considered each qubit as a two-level
bosonic mode and defined the operators $\hat{a}_{j},\hat{b}_{j},\hat{a}%
_{j}^{+},$ and $\hat{b}_{j}^{+}$ using the two energy eigenstates $%
\left\vert 0\right\rangle $ and $\left\vert 1\right\rangle $ as the
computational basis states. It is noted that during the operation, more than
a single photon could reside in each resonator when the large detuning
conditions (2) are not well satisfied. For this reason, we treat the
above-defined operators $\hat{a}_{j},\hat{b}_{j},\hat{a}_{j}^{+},$and $\hat{b%
}_{j}^{+}$ as the usual photon annihilation and creation operators
introduced in quantum optics. Note that after this replacement, the
Hamiltonian $H_{I}$ in the interaction picture, describing the interaction
of the four resonators with the transmon coupler, takes the same form as
that given in Eq. (1) with $N=2.$ By doing this, the effects of all excited
states of the resonators are taken into account.

The numerical simulation is carried out by solving the master equation (10)
which describes the dynamics of four resonators coupled to a superconducting
transmon. As shown in Table I [59-61,74-77], the simulation takes the
effects of dissipation and dephasing on the fidelity into account. \
Specifically,\ we selected a conservative set of resonator and transmon
parameters in the numerical simulation to demonstrate experimental
feasibility. In addition, assuming all coupling constants are equal\ $%
g_{1}=\mu _{1}=g_{2}=\mu _{2}\equiv g=2\pi \times 100$\ MHz (again this is
an undesirable situation).\textrm{\ }The fidelity of the operations is given
by $\mathcal{F}=\sqrt{\left\langle \psi _{id}\right\vert \widetilde{\rho }%
\left\vert \psi _{id}\right\rangle }$ [78], where $\left\vert \psi
_{id}\right\rangle =\left\vert \psi _{\mathrm{A}}\left( t\right)
\right\rangle \left\vert \psi _{\mathrm{B}}\left( t\right) \right\rangle
\left\vert g\right\rangle _{\mathrm{c}},$ with $\left\vert \psi _{\mathrm{A}%
}\left( t\right) \right\rangle $ given in Eq.~(8) and $\left\vert \psi _{%
\mathrm{B}}\left( t\right) \right\rangle $ in Eq.~(9), is the output state
for an ideal system (i.e., without dissipation, dephasing and leakage to
high excited states) after completing the operations and $\widetilde{\rho }$
is the final density operator of the system.

The simulated fidelity as a function of the\emph{\ }dimensionless\emph{\ }%
detuning $\alpha \equiv \Delta /g$ in the range of $4\leq \alpha \leq 10$
for Bell-state transfer and exchange are shown in Figs. 4 and 5,
respectively. It is found that the maximum fidelity of transferring the four
Bell states $\left\vert \psi ^{\pm }\right\rangle $ and $\left\vert \phi
^{\pm }\right\rangle $ from the resonators ($a_{1},a_{2}$) to ($b_{1},b_{2}$%
) or vice versa is equal to or better than $99.1\%,$ when $\alpha \equiv
\Delta /g=5.5$. While for exchanging $\left\vert \psi ^{+}\right\rangle $
with $\left\vert \psi ^{-}\right\rangle ,$ $\left\vert \phi
^{+}\right\rangle $ with $\left\vert \phi ^{-}\right\rangle ,$ $\left\vert
\phi ^{\pm }\right\rangle $ with $\left\vert \psi ^{+}\right\rangle $, and $%
\left\vert \phi ^{\pm }\right\rangle $ with $\left\vert \psi
^{-}\right\rangle $ the maximum fidelity is $97.2\%,$ $96.3\%,$ $96.4\%$,
and $96.6\%$, respectively, obtained around $\alpha =9.3$. \ Furthermore,
the high fidelity is hardly affected by weak residual inter-resonator
crosstalks as often the case in experimental situations (see Supplementary
Information). However, it should be pointed out that the value of the
detuning parameter $\alpha $ at which the maximum fidelity is achieved
depends on other parameters, such as the photon decay rate, of the
resonators and thus is not universal. \ In experiments, $\alpha $ needs to
be fine tuned to obtain the maximum fidelity.

As discussed previously, one of the advantages of the single-step TEAMS
method proposed here is that the coupler remains separable from the qubits
and it stays mostly in the ground state so that the effects of coupler's
decoherence on the fidelity of TEAMS is significantly reduced. \ To confirm
this property numerical simulations were performed and the result confirms
that for Bell-state transfer (exchange) the time-averaged population of the
coupler's excited state $|e\rangle $ is $0.03\leq $ $\overline{P}_{e}\leq
0.08$ ($0.03\leq $ $\overline{P}_{e}\leq 0.05$) for the operations described
above.

As the above example and parameters listed in Table 1 show, our scheme does
not require the use of tunable resonator-coupler coupling strength and/or
tunable frequency resonators. Furthermore, $g_{j}=\mu _{j}$\ is not a
necessary condition and it is chosen only to simplify discussions. The
strong condition that needs to be satisfied for simultaneous TEAMS is the
effective pair-wise coupling strength $\lambda _{j}=g_{j}\mu _{j}/\Delta _{j}
$\ should have the same value for all $j=1,2,..N$\ qubit pairs. \ Therefore,
our scheme does not require, though it would be more convenient, to have
tunable resonator-qubit coupling strength $g_{j}$\ and $\mu _{j}$. For
example, it is straightforward to design and to fabricate pairs of
resonators $a_{j}$ and $b_{j}$ to have $j$-dependent frequency $\omega _{j}$%
\ and coupling strength $g_{j}$\ such that $|\lambda _{j}|=g_{j}^{2}/|\Delta
_{j}|=\lambda .$\textrm{\ }

The advantage of utilizing positive as well as negative detunings is worth
to discuss. Because our scheme essentially explores the frequency
multiplexing property of the effective Hamiltonian (5) it will encounter the
"frequency crowding" problem. Because the system dynamics does not depend on
the signs of detunings according to Eqs. (6-8), utilizing the positive as
well as the negative detunings would double the maximum number of qubits
that can be accommodated by a given circuit. This advantage is most clearly
demonstrated by the example presented above: when all four resonators have
the same couplig strength to the coupler the only way to satisfy $\lambda
_{1}=|\lambda _{2}|=\lambda $\ is to have $\Delta _{1}=-\Delta _{2}$\textrm{%
. }

We would like to point out that\ although the proposed scheme of TEAMS can
be implemented using a small number of qubits or resonators with fixed
frequency and/or coupling strength it is in general diserable and even
necessary to have the frequency tunability for a moderate number of qubits
or resonators. This is especially true if one wants to realize the
reconfigurable network as that of illustrated in Fig. 1. Note that tunable
frequency artificial atoms are readily available and tunable superconducting
resonators have been demonstrated by incorporating nonlinear elements, such
as a small dc SQUID, into the design [57,58].\emph{\ }

\section*{Discussion}

We have shown that the minimum hardware resources required for
simultaneously transferring or swapping arbitrary multipartite quantum
states between two sets of otherwise noninteracting qubits each having a $%
2^N $-dimensional\ Hilbert space can be achieved using a single two-level
coupler. This result means that arbitrary $N$-qubit states that span a $2^N$%
-dimensional Hilbert space can be transferred or exchanged between two $N$%
-qubit registers in a single step via a coupler whose Hilbert space is $2$%
-dimensional only. In addition, during the entire process the coupler
remains separable from the qubits and stays mostly in the ground state\
throughout the entire process\ thus suppressing the undesirable effects of
coupler decoherence. The finding of the minimum resource required and the
method to simultaneously transfer or swap arbitrary $N$-partite states in a
single step is of great interest and fundamental importance in quantum
information science. If realized experimentally, it would be a big step
forward in the direction of building scalable quantum information processing
networks because in principle the operation time required is independent of
the number of qubits involved. In addition, as a concrete example we show
that transferring (exchanging) the Bell states between two pairs of
resonators (bosonic qubits) interacting via a superconducting transmon
coupler can achieve fidelity as high as $99.1\%$ (no less than $96.3\%$)
with conservative device and circuit parameters. \ In addition, because the
method does not use classical pulses during the entire operation and the
constituents of the two registers can be reassigned \textit{in situ} through
the reconfigurable coupler-mediated pair interaction described by Eq. (4)
and illustrated in Fig. 1(b), the proposed scheme can greatly reduce the
complexity of the circuit and can serve as one of the fundamental building
block for the development of more sophisticated quantum network
architectures in the future. Finally, the result presented here is general
and thus in principle can be applied to any type of physical qubits such as
electronic and nuclear spins, photons, atoms, and artificial atoms.

\section*{Methods}

\noindent \textbf{Master equation.} When the dissipation and dephasing are
included, the dynamics of the open system is determined by the following
master equation
\begin{eqnarray}
\frac{d\rho }{dt} &=&-i\left[ H_I,\rho \right] +\sum_{j=1}^2\kappa _{a_j}%
\mathcal{L}\left[ \hat a_j\right] +\sum_{j=1}^2\kappa _{b_j}\mathcal{L}\left[
\hat b_j\right]  \notag \\
&&\ +\gamma \mathcal{L}\left[ \sigma \right] \ +\gamma _\varphi \left(
\sigma _z\rho \sigma _z-\rho \right) ,
\end{eqnarray}
where $H_I$ is the interaction Hamiltonian given in Eq. (1), $\sigma
_z=\left| e\right\rangle \left\langle e\right| -\left| g\right\rangle
\left\langle g\right| ,$ and $\mathcal{L}\left[ \Lambda \right] =\Lambda
\rho \Lambda ^{+}-\Lambda ^{+}\Lambda \rho /2-\rho \Lambda ^{+}\Lambda /2$
(with $\Lambda =\hat a_j,\hat b_j,\sigma $). In addition, $\kappa _{a_j}$\ ($%
\kappa _{b_j}$) is the decay rate of the resonator mode $a_j$\ ($b_j$);\ $%
\gamma $\ is the energy relaxation rate for the level $\left| e\right\rangle
$; and $\gamma _\varphi $ is the dephasing rate of the level $\left|
e\right\rangle $ of the coupler\textbf{.}

\noindent \textbf{Effective Hamiltonian for non-identical coupling strengths
and detunings.} Suppose that qubit $a_j$ ($b_j$) is coupled to the coupler
C, with coupling strength $g_j$ ($\mu _j$) and detuning\ $\Delta _{a_j}$\ ($%
\Delta _{b_j}$). In the interaction picture, the Hamiltonian of the whole
system is given by
\begin{equation}
H_I=\sum_{j=1}^N\left( g_je^{i\Delta _{a_j}t}\hat a_j\sigma ^{+}+\mu
_je^{i\Delta _{b_j}t}\hat b_j\sigma ^{+}+\mathrm{H.c.}\right) ,
\end{equation}
where $\Delta _{a_j}=\omega _{\mathrm{c}}-\omega _{a_j}$ and $\Delta
_{b_j}=\omega _{\mathrm{c}}-\omega _{b_j}$.

Under the large detuning condition $\Delta _{a_j}\gg g_j$ and $\Delta
_{b_j}\gg \mu _j,$\ and when the detunings satisfy the following condition
\begin{equation}
\frac{\left| \Delta _{\alpha _j}-\Delta _{\beta _k}\right| }{\Delta _{\alpha
_j}^{-1}+\Delta _{\beta _k}^{-1}}\gg g_jg_k,\mu _j\mu _k,g_j\mu _k;j\neq k
\end{equation}
(where $\alpha _j\in \left\{ a_j,b_j\right\} $ and $\beta _k\in \left\{
a_k,b_k\right\} $), we can obtain the effective Hamiltonian $%
H_{eff}=H_0+H_{int},$ with
\begin{eqnarray}
H_0 &=&\sum_{j=1}^N\left( \frac{g_j^2}{\Delta _{a_j}}\hat a_j\hat
a_j^{\dagger }+\frac{\mu _j^2}{\Delta _{b_j}}\hat b_j\hat b_j^{\dagger
}\right) \left| e\right\rangle \left\langle e\right|  \notag \\
&&\ \ -\sum_{j=1}^N\left( \frac{g_j^2}{\Delta _{a_j}}\hat a_j^{\dagger }\hat
a_j+\frac{\mu _j^2}{\Delta _{b_j}}\hat b_j^{\dagger }\hat b_j\right) \left|
g\right\rangle \left\langle g\right| ,
\end{eqnarray}
\begin{equation}
H_{int}=\sum_{j=1}^N\lambda _j\left[ e^{i(\Delta _{a_j}-\Delta _{b_j})t}\hat
a_j\hat b_j^{\dagger }+\mathrm{H.c.}\right] (\left| e\right\rangle
\left\langle e\right| -\left| g\right\rangle \left\langle g\right| ),
\end{equation}
where $\lambda _j=\frac{g_j\mu _j}2(\Delta _{a_j}^{-1}+\Delta _{b_j}^{-1}).$%
\ When the coupler C is initially in the ground state $\left| g\right\rangle
$, it will remain in this state as the Hamiltonians $H_0$ and $H_{int}$\
cannot induce any transition for the coupler. In this case, the Hamiltonians
$H_0$ and $H_{int}$ reduce to
\begin{equation}
H_0=-\sum_{j=1}^N\left( \frac{g_j^2}{\Delta _{a_j}}\hat a_j^{\dagger }\hat
a_j+\frac{\mu _j^2}{\Delta _{b_j}}\hat b_j^{\dagger }\hat b_j\right) \left|
g\right\rangle \left\langle g\right| ,
\end{equation}
\begin{equation}
H_{int}=-\sum_{j=1}^N\lambda _j\left[ e^{i(\Delta _{a_j}-\Delta
_{b_j})t}\hat a_j\hat b_j^{\dagger }+\mathrm{H.c.}\right] \left|
g\right\rangle \left\langle g\right| ,
\end{equation}

In a new interaction picture\textrm{\ }with respect to the Hamiltonian $H_0,$
we obtain
\begin{eqnarray}
H_{int}^{\prime } &=&e^{iH_0t}H_{int}e^{-iH_0t}  \notag \\
\ &=&-\sum_{j=1}^N\lambda _j\left[ e^{i(g_j^2/\Delta _{a_j}-\mu _j^2/\Delta
_{b_j})t}e^{i(\Delta _{a_j}-\Delta _{b_j})t}\hat a_j\hat b_j^{\dagger }+%
\mathrm{H.c.}\right] \left| g\right\rangle \left\langle g\right| .
\end{eqnarray}
For the setting
\begin{equation}
g_j^2/\Delta _{a_j}-\mu _j^2/\Delta _{b_j}=-(\Delta _{a_j}-\Delta _{b_j}),
\end{equation}
the Hamiltonian (17) becomes
\begin{equation}
H_{int}^{\prime }=-\sum_{j=1}^N\lambda _j(\hat a_j\hat b_j^{\dagger }+\hat
a_j^{\dagger }\hat b_j)\left| g\right\rangle \left\langle g\right| ,
\end{equation}
which is exactly the one given in Eq. (5) after dropping the atomic operator
$\left| g\right\rangle \left\langle g\right| $.

Note that condition (18) can be achieved by setting
\begin{equation}
\Delta _{b_j}=\frac{\Delta _{a_j}^2+g_j^2+\sqrt{(\Delta
_{a_j}^2+g_j^2)^2-4\Delta _{a_j}^2\mu _j^2}}{2\Delta _{a_j}}.
\end{equation}
For $g_j=\mu _j,$ we have $\Delta _{b_j}=\Delta _{a_j}$, i.e., the case that
we discussed previously. In constrast, for $g_j\neq \mu _j,$ we have $\Delta
_{b_j}\neq \Delta _{a_j}$ from Eq. (20). This result implies that if the
coupling $g_j$ is not equalivent to $\mu _j$, one can still obtain the
\textit{time-independent} effective Hamiltonian (5) or (19) by setting the
detuning $\Delta _{b_j}$ slightly different from $\Delta _{a_j}$.

\begin{addendum}

\item[Acknowledgments]

S. Han acknowledges support from NSF of the United States (PHY-1314861)
and partial support by DMEA. S.B. Zheng was supported by the Major State Basic Research Development Program of
China under Grant No. 2012CB921601. Q.P.S. was supported by the National
Natural Science Foundation of China under Grant No. 11247008. C.P.Y. was
supported in part by the National Natural Science Foundation of China under
Grant Nos. 11074062 and 11374083, the Zhejiang Natural Science Foundation
under Grant No. LZ13A040002, and the funds from Hangzhou Normal University
under Grant Nos. HSQK0081 and PD13002004. This work was also supported by
the funds from Hangzhou City for the Hangzhou-City Quantum Information and
Quantum Optics Innovation Research Team.

\item[Author contributions]

C.P.Y., S.B.Z. and S.H. carried out analytical calculations and wrote the
main manuscript text. Q.P.S. performed numerical simulation. All authors
participated in discussing the results and contributed to the interpretation
of the work.

\item[Additional information]
Competing financial interests: The authors declare no competing financial
interests.
\end{addendum}


\clearpage

\textbf{Table 1:} \textbf{Parameters for a transmon-coupled multi-resonator
system}. The values of $\omega _{a_{j}},$ $\omega _{b_{j}},$ $Q_{a_{j}},$
and $Q_{b_{j}}$ ($j=1,2$) are estimated for $\alpha =5.5$ (Bell-state
transfer), $\alpha =9.3$ (Bell-state exchange), $\omega _{c}/2\pi =6.5$ GHz,
and $g/2\pi =100$ MHz. Here, $Q_{a_{j}}=\omega _{a_{j}}\kappa _{a_{j}}^{-1}$
and $Q_{b_{j}}=\omega _{b_{j}}\kappa _{b_{j}}^{-1}$. $T_{1}$ and $T_{2}$ can
be made to be on the order of $20-60$ $\mu $s for state-of-the-art
superconducting transom devices [74-76]. Superconducting CPW (coplanar
waveguide) resonators with a quality factor $Q\sim 10^{6}$ have been
experimentally demonstrated [59-61]. In addition, the coupling strength $%
g/2\pi \sim 360$\textbf{\ }MHz has been reported for a superconducting
transmon qubit coupled to a one-dimensional standing-wave CPW resonator [77].

\textbf{Figure 1:} \textbf{Two sets of qubits coupled by a two-level coupler
C}. Here, the large circle at the center represents the two-level coupler C,
the smaller circles on the left (right) indicate the $N$ qubits $%
a_{1},a_{2},...,a_{N}$ ($b_{1},b_{2},...,b_{N}$) in the register A (B)
connected to the coupler C by lines with the same color form an interacting
qubit pair. In (a), the $N$ pairs of qubits are ($a_{1},b_{1}$), ($%
a_{2},b_{2}$),..., and ($a_{N},b_{N}$); while in (b) the $N$ pairs of qubits
are randomly chosen as, e.g., ($a_{1},b_{2}$), ($a_{2},b_{N}$),..., and ($%
a_{N},b_{1}$). For (a) and (b), arbitrary $N$-partite states can be
transferred or exchanged between A and B. In addition, various entangled
states of qubits in A and B can be generated by the same coupler mediated
qubit-qubit interaction. \bigskip

\textbf{Figure 2:} \textbf{Illustration of qubit-coupler dispersive
interaction}. The two horizontal solid lines represent the two energy levels
of the coupler C. The bottom dashed line represents the common ground energy
level of the $2N$ qubits, while the top dashed lines in different colors
represent the higher energy levels of the $2N$ qubits, respectively. A
vertical line, linked to the bottom dashed line and a top dashed line,
represents the level spacing between the two energy levels of a qubit. The
frequency of qubit $a_{j}$ ($b_{j}$) is labelled as $\omega _{aj}$ ($\omega
_{bj}$) (not shown), while the frequency of the coupler C is denoted as $%
\omega _{c}$ (not shown). Qubit $a_{j}$ ($b_{j}$) is dispersively coupled to
the coupler C with coupling constant $g_j$ ($\mu _j$) and detuning $\Delta_j$
($j=1,2,...,N$). Here, $\Delta _{j}=\omega _{\mathrm{c}}-\omega _{aj}=\omega
_{\mathrm{c}}-\omega _{bj}$. \bigskip

\textbf{Figure 3:} \textbf{Setup for four resonators $%
a_{1},a_{2},b_{1},b_{2} $ coupled by a superconducting transmon coupler
(i.e., the circle C)}. Each resonator here is a one-dimensional coplanar
waveguide resonator. The superconducting transmon qubit is capacitively
coupled to each resonator via a capacitance. \bigskip

\textbf{Figure 4:} \textbf{Fidelity versus $\alpha $ for the Bell-state
transfer}. Here, the red and blue curves correspond to transferring the two
Bell states $\left\vert \psi ^{+}\right\rangle $ and $\left\vert \psi
^{-}\right\rangle $, respectively. Numerical simulation shows that the
fidelity for transferring the other two Bell states $\left\vert \phi ^{\pm
}\right\rangle $ is the same (the green line). \bigskip

\textbf{Figure 5:} \textbf{Fidelity versus $\alpha $ for the Bell-state
exchange}. Here, the red, blue, green, and yellow curves correspond to
exchanging the Bell states, $\left\vert \psi ^{+}\right\rangle $ with $%
\left\vert \psi ^{-}\right\rangle ,$ $\left\vert \phi ^{+}\right\rangle $
with $\left\vert \phi ^{-}\right\rangle ,$ $\left\vert \phi ^{\pm
}\right\rangle $ with $\left\vert \psi ^{+}\right\rangle $, and $\left\vert
\phi ^{\pm }\right\rangle $ with $\left\vert \psi ^{-}\right\rangle $,
between the qubit pairs ($a_{1},a_{2}$) and ($b_{1},b_{2}$), respectively.
\bigskip

\begin{center}
\begin{table}[tbp]
Table 1\newline
\newline
\begin{tabular}{|c|c|c|c|}
\hline
Parameter & Symbol & Bell-state exchange & Bell-state transfer \\ \hline
Resonator photon lifetime & $\kappa _{a_1}^{-1},\kappa _{b_1}^{-1},\kappa
_{a_2}^{-1},\kappa _{b_2}^{-1}$ & $1$ $\mu $s & $1$ $\mu $s \\ \hline
Coupler energy relaxation time & $\gamma ^{-1}$ & $3$ $\mu $s & $3$ $\mu $s
\\ \hline
Coupler dephasing time & $\gamma _\varphi ^{-1}$ & $3$ $\mu $s & $3$ $\mu $s
\\ \hline
Coupler frequency & $\omega _c/2\pi $ & $6.0$ GHz & $6.0$ GHz \\ \hline
Resonator frequency, pair I & $\omega _{a_1}/2\pi ,$ $\omega _{b_1}/2\pi $ &
$5.07$ GHz & $5.45$ GHz \\ \hline
Resonator frequency, pair II & $\omega _{a_2}/2\pi ,$ $\omega _{b_2}/2\pi $
& $6.93$ GHz & $6.55$ GHz \\ \hline
Resonator quality factor, pair I & $Q_{a_1},$ $Q_{b_1}$ & $3.2\times 10^4$ &
$3.4\times 10^4$ \\ \hline
Resonator quality factor, pair II & $Q_{a_2},$ $Q_{b_2}$ & $4.4\times 10^4$
& $4.1\times 10^4$ \\ \hline
\end{tabular}%
\end{table}
\end{center}

\clearpage
Figure 1
\begin{figure}
\begin{center}
\epsfig{file=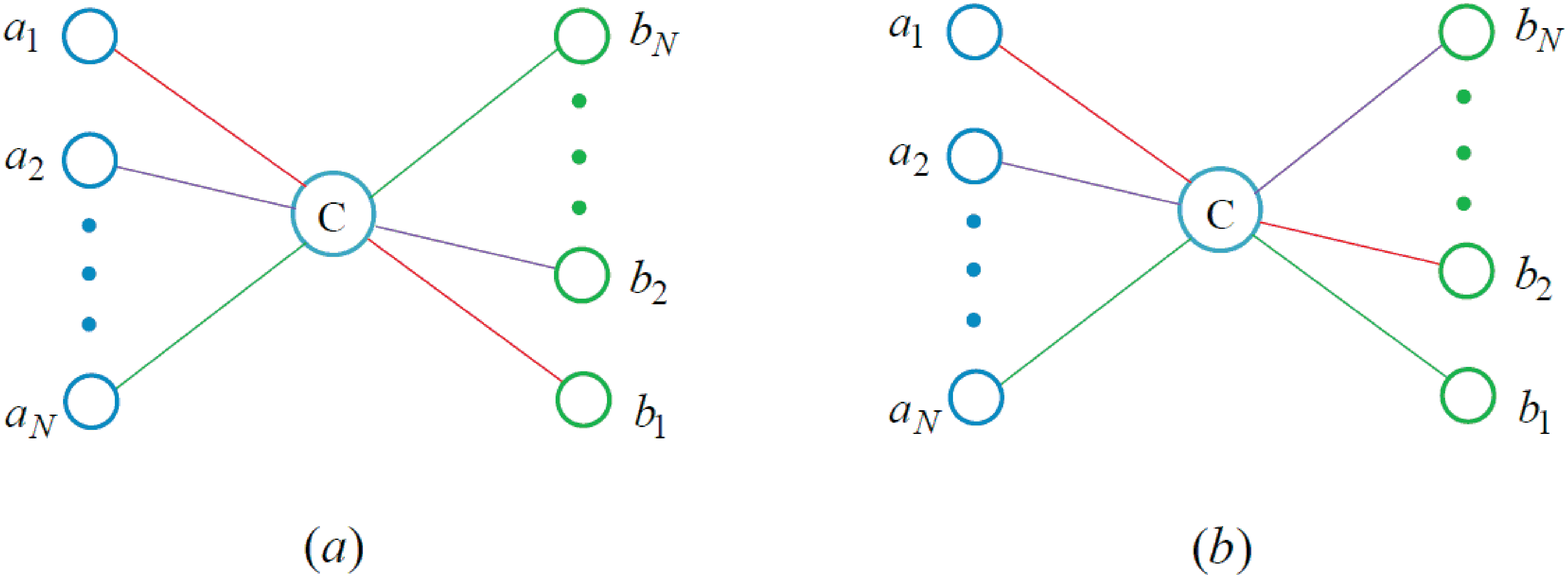,width=15cm}
\end{center}
\label{fig:1}
\end{figure}
\clearpage
Figure 2
\begin{figure}
\begin{center}
\epsfig{file=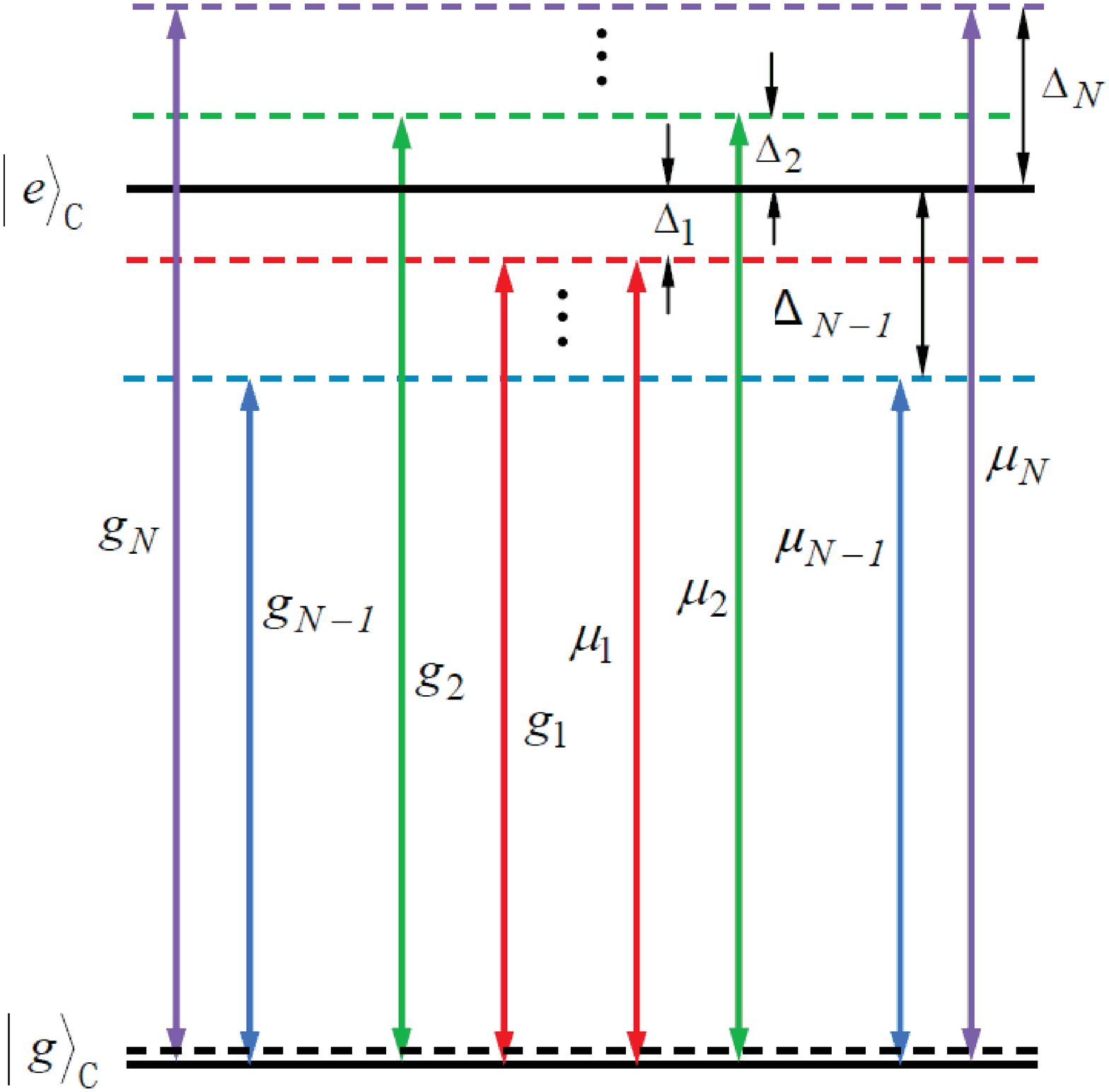,width=15cm}
\end{center}
\label{fig:2}
\end{figure}
\clearpage
Figure 3
\begin{figure}
\begin{center}
\epsfig{file=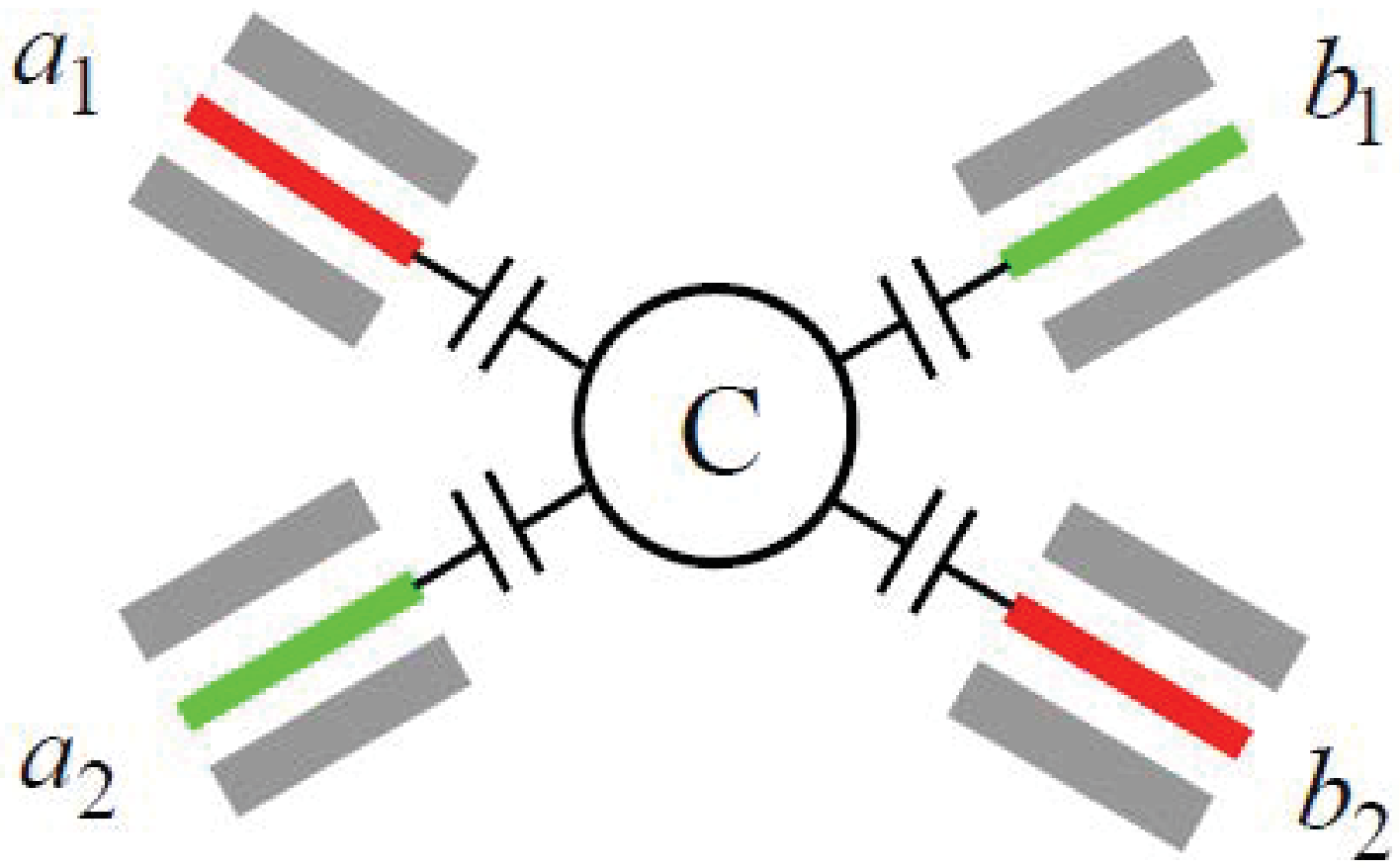,width=15cm}
\end{center}
\label{fig:3}
\end{figure}
\clearpage
Figure 4
\begin{figure}
\begin{center}
\epsfig{file=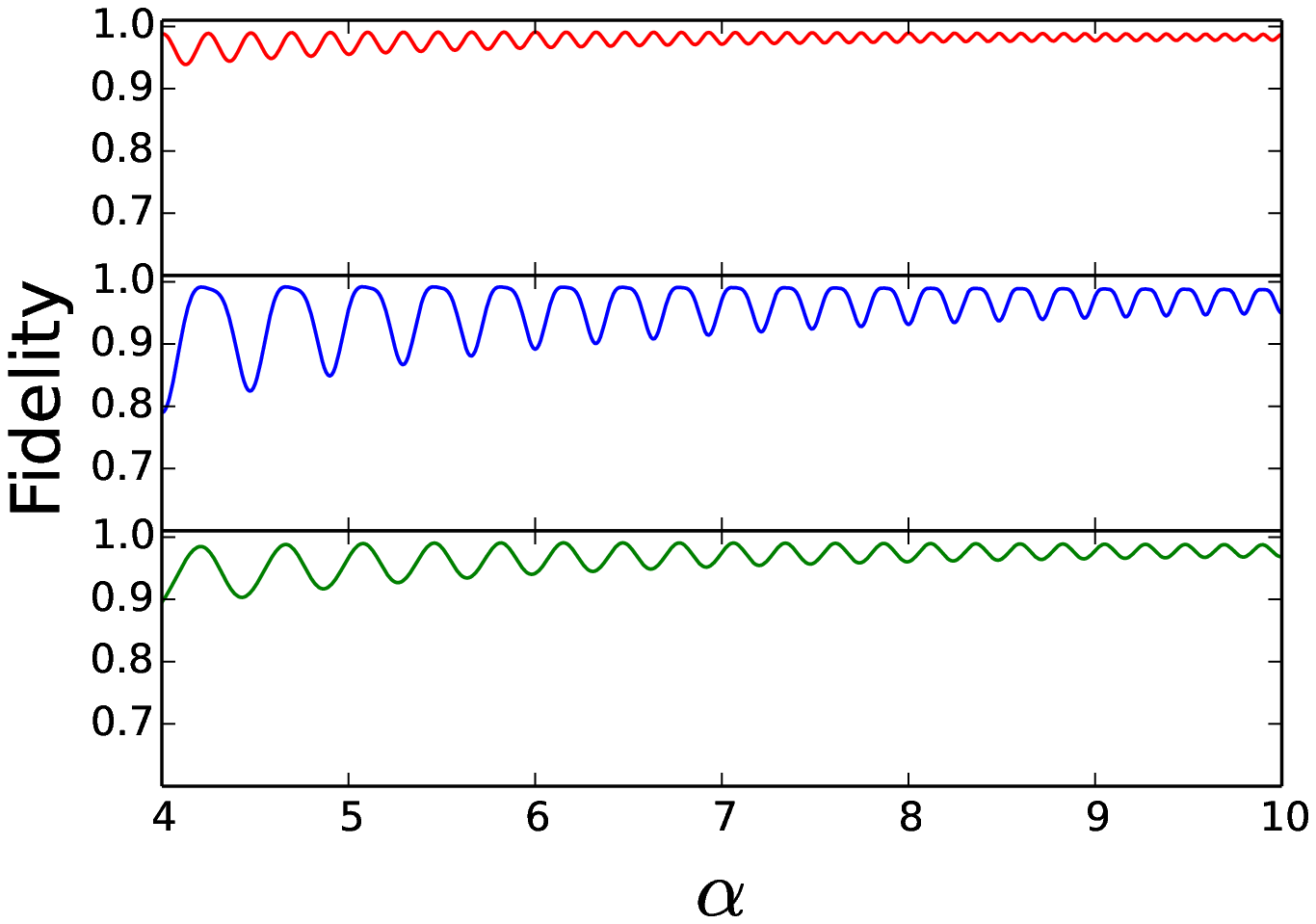,width=15cm}
\end{center}
\label{fig:4}
\end{figure}
\clearpage
Figure 5
\begin{figure}
\begin{center}
\epsfig{file=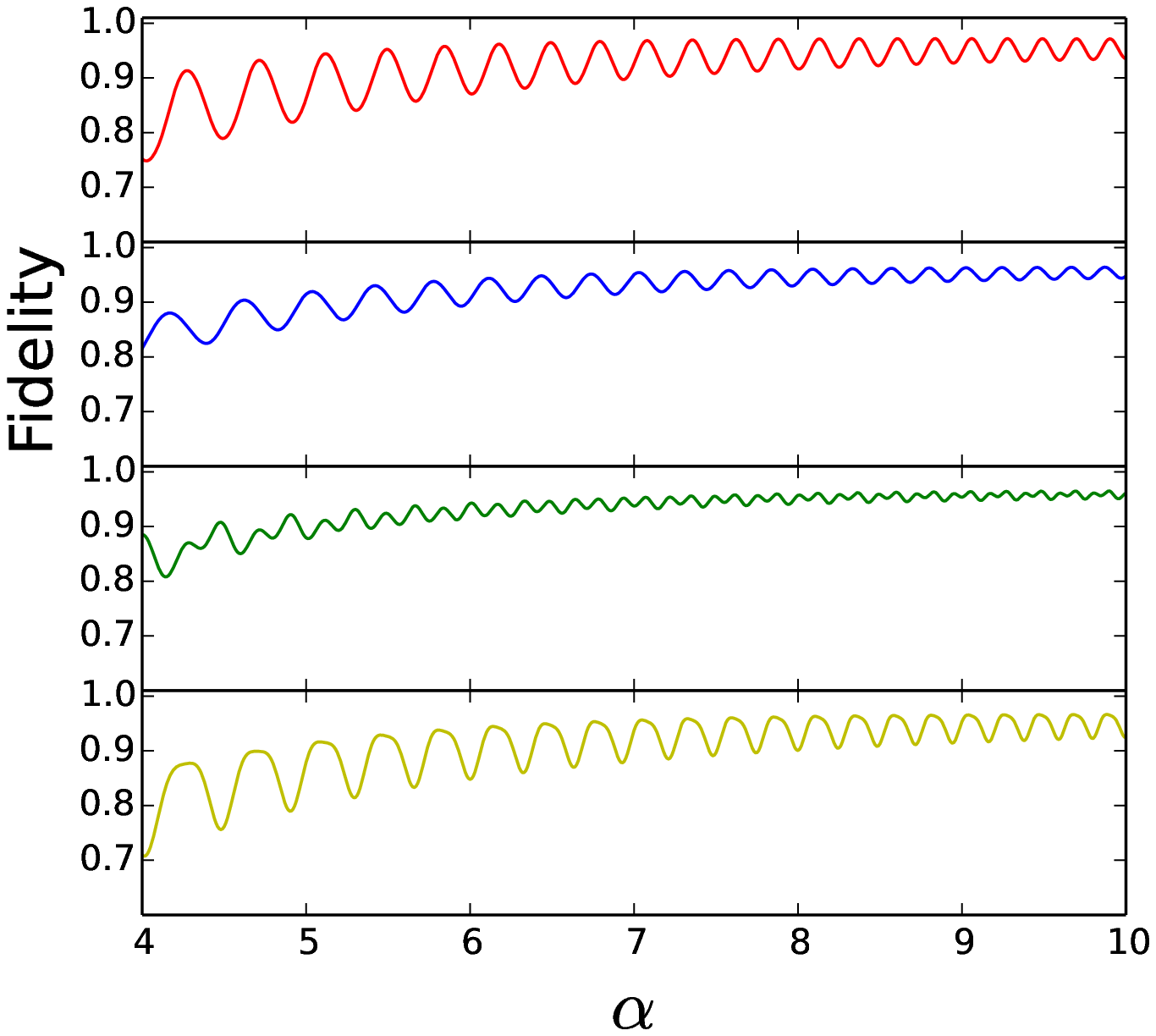,width=15cm}
\end{center}
\label{fig:5}
\end{figure}

\clearpage

\noindent\textbf{{\Large \sffamily{Supplementary Information}}}

\renewcommand\thefigure{S\arabic{figure}}
\setcounter{figure}{0}
\renewcommand\theequation{S\arabic{equation}}
\setcounter{equation}{0} \renewcommand\thepage {S\arabic{page}} %
\setcounter{page}{1}
\renewcommand\thetable{S\arabic{table}}
\setcounter{table}{0}

\bigskip

When the inter-cavity crosstalk between resonators are considered, the
Hamiltonian (1) is modified as follows
\begin{eqnarray}
H_{\mathrm{I}}^{\prime } &=&\sum_{j=1}^2\left( g_je^{i\Delta _jt}\hat
a_j\sigma ^{+}+\mu _je^{i\Delta _jt}\hat b_j\sigma ^{+}+\mathrm{H.c.}\right)
\nonumber \\
&&\ \ \ +\left( g_{a_1a_2}e^{i\delta t}a_1a_2^{\dagger
}+g_{a_1b_2}e^{i\delta t}a_1b_2^{\dagger }+\mathrm{H.c.}\right)  \nonumber \\
&&\ \ \ +\left( g_{a_2b_1}e^{-i\delta t}a_2b_1^{\dagger
}+g_{b_2b_1}e^{-i\delta t}b_2b_1^{\dagger }+\mathrm{H.c.}\right)  \nonumber
\\
&&\ \ \ +\left( g_{a_1b_1}a_1b_1^{\dagger }+g_{a_2b_2}a_2b_2^{\dagger }+%
\mathrm{H.c.}\right) ,
\end{eqnarray}
where the terms in the last three lines represent the inter-cavity crosstalk
between any two resonators, with the coupling constants ($%
g_{a_1a_2},g_{a_1b_2},g_{a_2b_1},g_{b_2b_1},g_{a_1b_1},g_{a_2b_2}$) and
detuning $\delta =\omega _{a_2}-\omega _{a_1}=\omega _{b_2}-\omega
_{a_1}=\omega _{a_2}-\omega _{b_1}=\omega _{b_2}-\omega _{b_1}$ of the two
associated resonators, due to $\omega _{a_1}=\omega _{b_1}$ and $\omega
_{a_2}=\omega _{b_2}$.

The numerical simulation is performed by solving the master equation (10),
with the Hamiltonian $H_{\mathrm{I}}$ there replaced by $H_{\mathrm{I}%
}^{\prime }.$ For simplicity, we set $%
g_{a_1a_2}=g_{a_1b_2}=g_{a_2b_1}=g_{b_2b_1}=g_{a_1b_1}=g_{a_2b_2}\equiv 0.01g
$ (a conservative consideration for weak direct inter-resonator crosstalks).
In our numerical simulation, the detuning setting $\Delta _1=-\Delta
_2=\Delta ,$ the coupler-resonator coupling constants $g_1=\mu _1=g_2=\mu
_2=g=2\pi \times 100$ MHz, the resonator photon lifetime, and the
decoherence time of the coupler are the same as those used for Figs. (4) and
(5) of the main text. The operational fidelity as a function of the
dimensionless detuning $\alpha \equiv \Delta /g$ in the range of $4\leq
\alpha \leq 10$ for Bell state transfer and exchange are plotted in Figs. S1
and S2, respectively. Compared Fig. S1 (S2) with Fig. 4 (5) of the main tex,
it can be seen that the high fidelity is hardly affected by weak direct
inter-resonator crosstalks for both Bell state transfer and exchange.



\clearpage
\textbf{Figure S1:} \textbf{Fidelity versus $\alpha $ for the Bell-state transfer.} The
curves in (a), (b), and (c) correspond to transferring the Bell states $%
\left| \psi ^{+}\right\rangle ,$ $\left| \psi ^{-}\right\rangle $, and $%
\left| \phi ^{\pm}\right\rangle $, respectively. Here, the red curves are
plotted without considering the inter-resonator crosstalks, while the blue
ones take the weak inter-resonator crosstalks into account. \bigskip

\textbf{Figure S2:} \textbf{Fidelity versus $\alpha $ for the Bell-state exchange}. The
curves in (a), (b), (c), and (d) correspond to exchanging the Bell states,
i.e, $\left| \psi ^{+}\right\rangle $ with $\left| \psi ^{-}\right\rangle ,$
$\left| \phi ^{+}\right\rangle $ with $\left| \phi ^{-}\right\rangle ,$ $%
\left| \phi ^{\pm }\right\rangle $ with $\left| \psi ^{+}\right\rangle $,
and $\left| \phi ^{\pm }\right\rangle $ with $\left| \psi ^{-}\right\rangle $%
, respectively. Here, the red curves are plotted without considering the
inter-resonator crosstalks, while the blue ones are plotted by taking the
weak inter-resonator crosstalks into account. \bigskip

\clearpage
Figure S1
\begin{figure}
\begin{center}
\epsfig{file=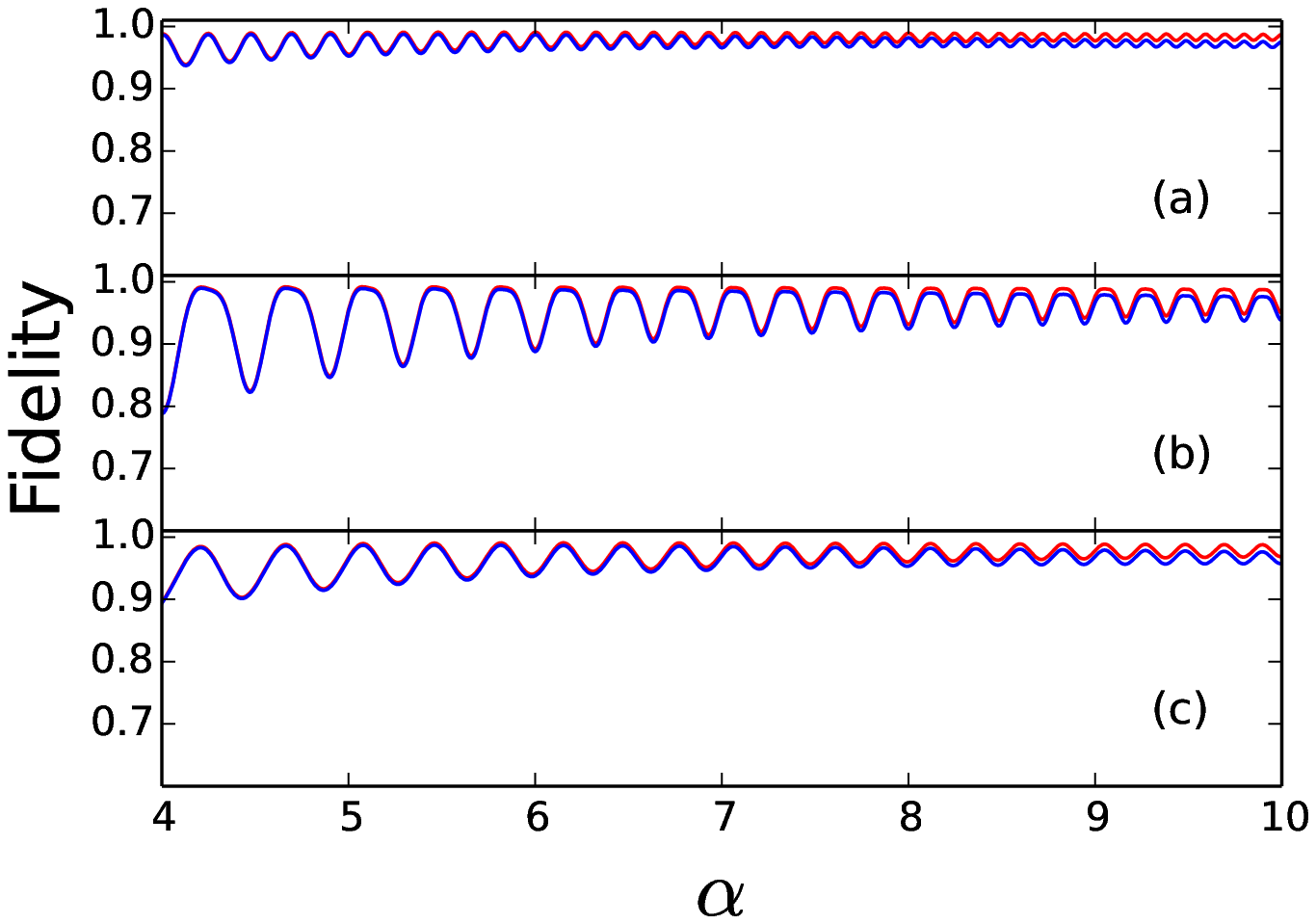,width=15cm}
\end{center}
\label{figS1}
\end{figure}

\clearpage
Figure S2
\begin{figure}
\begin{center}
\epsfig{file=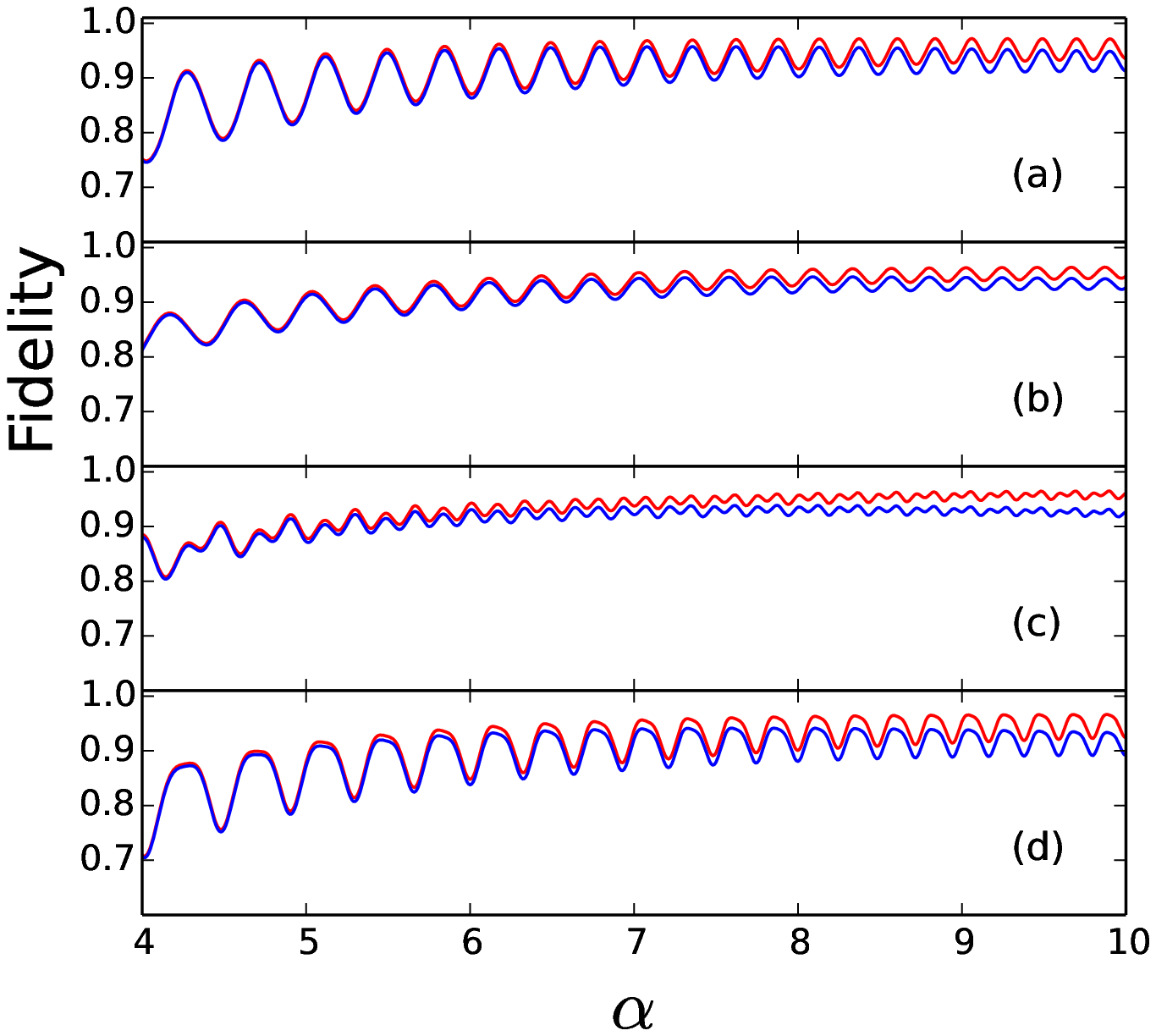,width=15cm}
\end{center}
\label{figS2}
\end{figure}

\end{document}